\documentclass[11pt]{article}

\pdfoutput=1

\usepackage{graphicx}
\usepackage[small]{caption}

\newcommand\tf{t_{\!f}}
\newcommand\tp{t_p}

\def\urldot{.\discretionary{}{}{}}
\def\urlslash{/\discretionary{}{}{}}

\begin{document}

\title{{\bf Predicting future duration from present age:\\
Revisiting a critical assessment of Gott's rule}}

\author{{\Large Carlton M.~Caves}
\\
\\
Department of Physics and Astronomy, MSC07--4220, University of New Mexico,\\
Albuquerque, New Mexico 87131-0001, USA\\
\\
and\\
\\
Department of Physics, University of Queensland,\\
Brisbane, Queensland 4072, Australia
\\
\\
E-mail: caves@info.phys.unm.edu
}

\date{2008 June~21}

\maketitle

\begin{abstract}
Gott has promulgated a rule for making probabilistic predictions of
the future duration of a phenomenon based on the phenomenon's present
age [{\sl Nature\/} {\bf 363}, 315 (1993)].  I show that the two
usual methods for deriving Gott's rule are flawed.  Nothing licenses
indiscriminate use of Gott's rule as a predictor of future duration.
It should only be used when the phenomenon in question has no
identifiable time scales.
\end{abstract}

\baselineskip=14.2pt

\section{Introduction}

In an article$^1$ published in {\sl Nature\/} in 1993 and in
subsequent publications$^{\hbox{\scriptsize2--5}}$ and the concluding
chapter of a book,$^6$ J.~Richard Gott~III has promulgated a formula
for making probabilistic predictions of the future duration of a
phenomenon based on the phenomenon's present age.  When you observe
that a phenomenon has lasted a time $\tp$, Gott instructs you to
predict that the phenomenon will last an additional time $\tf\ge
Y\tp$ with probability
\begin{equation}
G(\tf\ge Y\tp)={1\over1+Y}\;.
\label{eq:Grule}
\end{equation}
For example, Gott's rule predicts that a phenomenon has a probability
of $1/2$ to survive an additional time at least as long as its
past ($Y=1$) and, by the same token, a probability of $1/2$ to end
before reaching twice its present age.

In applying his rule to a host of phenomena, Gott usually couches his
predictions in terms of a particular 95\% confidence interval, 95\%
confidence being his standard for a scientific prediction.  According
to his rule, the probability that a phenomenon's future duration,
$\tf$, will be between $1/39$ and $39$ times its present age, $\tp$,
is $G(\tf\ge\tp/39)-G(\tf\ge39\tp)=39/40-1/40=0.95$. The flip side of
this prediction is that the phenomenon has a 2.5\% chance to end
before reaching 1/39 of its present age and a 2.5\% chance of lasting
longer than 39 times its present age.

Gott bases his formula on a temporal version of the Copernican
principle: when you observe the phenomenon's present age, your
observation does not occur at a special time.  Here I show,
distilling the essence of a previous critical analysis$^7$ of Gott's
work, that although the Copernican principle does lead directly to a
version of Gott's rule, this version is essentially meaningless and,
in particular, does not authorize his predictions for future
longevity based on present age.

In published papers and his book, Gott is on record as applying his
rule to
himself,$^{2,6}$
Christ\-ianity,$^6$
the former Soviet Union,$^{1,6}$
the Third Reich,$^6$
the United States,$^6$
Canada,$^4$
world leaders,$^{2,4,6}$
Stonehenge,$^4$
the Seven Wonders of the Ancient World,$^6$
the Pantheon,$^6$
the Great Wall of China,$^6$
{\sl Nature},$^1$
the {\sl Wall Street Journal},$^6$
{\sl The New York Times},$^6$
the Berlin Wall,$^{\hbox{\scriptsize1--4,6}}$
the Astronomical Society of the Pacific,$^2$
the 44 Broadway and off-Broadway plays open
and running on 27~May 1993,$^{\hbox{\scriptsize2--4,6}}$
the Thatcher-Major Conservative government in the UK,$^{\hbox{\scriptsize2--4,6}}$
Manhattan (New York City),$^6$
the New York Stock Exchange,$^6$
Oxford University,$^6$
the internet,$^6$
Microsoft,$^6$
General Motors,$^6$
the human spaceflight program,$^{\hbox{\scriptsize1--6}}$
and
{\it homo sapiens}.$^{\hbox{\scriptsize1--6}}$
In all these cases---even the New York plays---Gott uses his rule to
make probabilistic predictions for the survival of individual
phenomena whose present age is known. For example, given {\sl
Nature\/}'s 123 years of publishing in 1993, Gott predicted that {\sl
Nature\/} had a 95\% chance to continue publishing for a period
between 3.15 years (already exceeded) and 4,800 years.$^1$  Most
notably, Gott has used the 200,000-year present age of {\it homo
sapiens} to predict that we have a 95\% chance to go extinct sometime
between 5,100 years and 7.8 million years from
now.$^{\hbox{\scriptsize1--6}}$  Although Gott issues occasional
cautionary statements about the applicability of his rule,$^{2,6}$
the list of phenomena to which he has applied the rule indicates that
these cautions don't cramp his style much.

Gott's predictions have received attention in the popular media,
including a favorable piece by Timothy Ferris in {\sl The New
Yorker},$^8$ which highlighted Gott's predictions for human survival
(and which motivated me to write my original paper$^7$), and a recent
article by John Tierney in {\sl The New York Times},$^9$ which
focused on the implications for space colonization.  My late 1999
posting of the paper$^7$ that eventually appeared in {\sl
Contemporary Physics\/} prompted a sympathetic article in {\sl The
New York Times\/} by James Glanz,$^{10}$ then a science writer at
{\sl The Times}.

Glanz, as a long-suffering fan of the Chicago White Sox, was
particularly interested in Gott's prediction, issued in 1996, for the
Sox's World Series prospects.  Gott opined that the Sox, having not
won a World Series title since 1917, would, with 95\% confidence, win
a Series sometime between 1999 and 5077.  In his 2000 article, Glanz
noted that the Sox hadn't yet succeeded, but he was clearly dismayed
by the long wait evoked by the mere mention of 5077.  Happily for him
and other Sox fans, they did win the Series in 2005.  In 1996 Gott
would have predicted a World Series title in 2005 or before with
probability~0.10, considerably less than the probability,
$1-(29/30)^9=0.26$, that comes from assuming that the Sox had the
same chance each year as the other 30 major-league ball clubs.

Gott has given two main derivations of his rule: the argument from
the Copernican principle, which he calls the delta-$t$
argument,$^{\hbox{\scriptsize1--6}}$ and a Bayesian
analysis,$^{2,6,11}$ which he adopted from criticism due to
Buch.$^{12}$  Both of these derivations are flawed.$^7$  Here I begin
in Sec.~\ref{sec:deltat} with an analysis of the delta-$t$ argument,
because it led Gott to his rule and because he consistently portrays
it as the chief justification for his predictions.  I show that the
delta-$t$ argument does not lead to any prediction of future duration
based on present age.  I then turn in Sec.~\ref{sec:Bayesian} to the
usual Bayesian derivation of Gott's rule, which has greater appeal to
most other contributors to the literature on the subject.  I
demonstrate that this derivation is simply wrong and sketch the
correct Bayesian analysis.  A concluding Sec.~\ref{sec:conclusion}
considers the assumptions that are required to get Gott's
predictions.

It should be emphasized at the outset that we will not conclude that
Gott's rule is ``wrong,'' but rather that its two primary derivations
are wrong.  In science flawed justifications are as bad as---perhaps
worse than---being obviously wrong, because they are more pernicious.
They can mislead you into using methods that don't apply in your
situation and can get you into trouble when you export those methods
to other contexts.  Determining the assumptions that underlie
whatever you are doing in science is essential, so that you know when
to abandon what you are doing in favor of something else.  The
purpose of this article is thus to debunk the two primary derivations
of Gott's rule and to identify the assumptions that underlie Gott's
rule in its predictive form, so that you will know what you are doing
should you choose to use it.

The discussion in this paper is couched mainly in terms of a simple
graphical representation, which is equivalent to the more formal,
Bayesian analysis given in Ref.~7.  Section~\ref{sec:deltat} on the
delta-$t$ argument is phrased almost entirely in terms of the
graphical representation.  The results of the Bayesian analysis in
Sec.~\ref{sec:Bayesian} can be understood by referring to the
graphical representation, but the Bayesian equations are included for
those who prefer to see the details.

The evidence from papers$^{\hbox{\scriptsize13--17}}$ that cite
Ref.~7 is that its argument and conclusions have not been appreciated
and understood. The goal of the present paper, with its graphical
mode of presentation, is to rectify that situation.

\section{Copernican ensembles and the delta-$t$ argument}
\label{sec:deltat}

The delta-$t$ argument is short and sweet.  It starts from the
premise that if your observation does not occur at a special
time---that is the temporal Copernican principle---then it is equally
likely to occur at any time within the total duration $T=\tp+\tf$.
This means that the probability that the present age $\tp$ is less
than or equal to $XT$, where $X$ is between 0 and 1 inclusive, is
$G(\tp\le XT)=X$. This being the same as the probability that the
future duration $\tf$ is not smaller than $(1-X)T$, i.e., not smaller
than $(X^{-1}-1)\tp$, one obtains Gott's rule~(\ref{eq:Grule}) by
letting $Y=X^{-1}-1$.

The alluring simplicity of the delta-$t$ argument means that we need
an equally simple way of investigating its validity and
interpretation.  In any probabilistic analysis, you start with a
prior probability density, in this case a distribution $w(T)$, which
gives the probability $w(T)\,dT$ that the phenomenon's total duration
lies in the interval between $T$ and $T+dT$.  This prior probability
density is based on whatever information or data you have about the
phenomenon before observing it.  To formulate the problem in terms of
the temporal Copernican principle, the description in terms of
duration $T$ must be supplemented by introducing an additional
temporal variable.  In doing so, it is convenient to set the
arbitrary zero of time at the present, i.e., at the time you observe
the phenomenon, and to let $t_0$ denote the time when the phenomenon
starts.  With these choices, the present age is $\tp=-t_0$, and the
future duration is $\tf=T+t_0$. The phenomenon is now characterized
by two variables, either $t_0$ and $T$ or $\tp$ and $\tf$.

The temporal Copernican principle---that you are not at a special
time relative to the phe\-no\-menon---is implemented by saying that
all starting times are equally likely, independent of duration~$T$.
More precisely, one requires that the joint probability density be
invariant under time translations,$^{18}$ which yields the unique
probability density
\begin{equation}
p(\tp,\tf)=p(t_0,T)=\gamma w(T)\;,
\end{equation}
where $\gamma$ is a constant, describing a uniform distribution for
the starting time.  That this distribution cannot be normalized turns
out not to be a problem, but it can be dealt with at this stage, if
desired, by cutting off the distribution at very large negative and
positive values of~$t_0$.

\begin{figure}[t]
\begin{center}
\includegraphics[height=12cm]{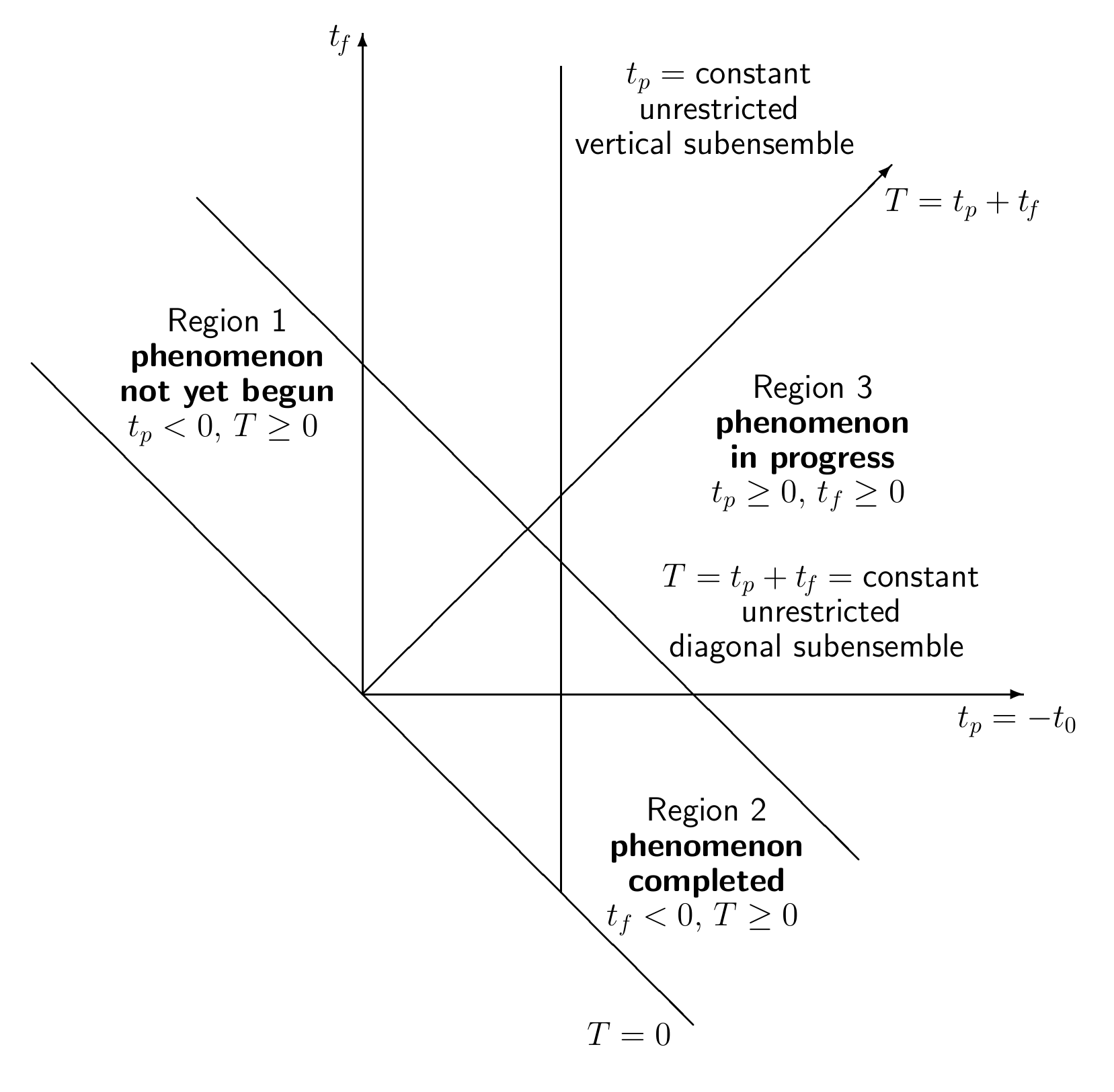}
\end{center}
\vspace{-24pt}
\caption{The $\tp$-$\tf$ plane on which the {\it unrestricted
Copernican ensemble\/} resides.
\label{fig1}}
\end{figure}

It is instructive to think about the joint probability density in
terms of an ensemble made up of many instances of the same
phenomenon.  We can picture this ensemble, which I call an {\it
unrestricted Copernican ensemble}, as a population distributed in a
plane whose horizontal axis is labeled by $\tp$ and whose vertical
axis is labeled by $\tf$.  The population density is proportional to
the probability density $p(\tp,\tf)=\gamma w(t_p+t_f)$.  The
Copernican plane is depicted in Fig.~\ref{fig1}.

The duration $T=\tp+\tf$ labels an axis that points symmetrically
into the first quadrant.  The constraint of nonnegative durations
means that the ensemble occupies the upper right half-plane, which
splits naturally into three regions:
\begin{enumerate}
\item{The upper left wedge ($\tp=-t_0<0$), in which the phenomenon
has not yet begun.}
\item{The lower right wedge ($\tf<0$), in which the phenomenon is over.}
\item{The first quadrant ($\tp\ge0$, $\tf\ge0$), in which the phenomenon is
in progress.}
\end{enumerate}
There are two instructive ways of dividing the unrestricted
Copernican ensemble into subensembles.  First, for each starting time
$t_0=-\tp$, the population along the associated vertical line is the
subensemble of durations for a phenomenon that starts at $t_0$. The
translational symmetry of the Copernican ensemble means that all
these {\it unrestricted vertical subensembles\/} describe the same
distribution of durations, given by the prior density $w(T)$. Second,
population is distributed uniformly along the diagonal lines of
constant duration $T$, each of which can be called an {\it
unrestricted diagonal subensemble}.

\begin{figure}[t]
\begin{center}
\includegraphics[height=10cm]{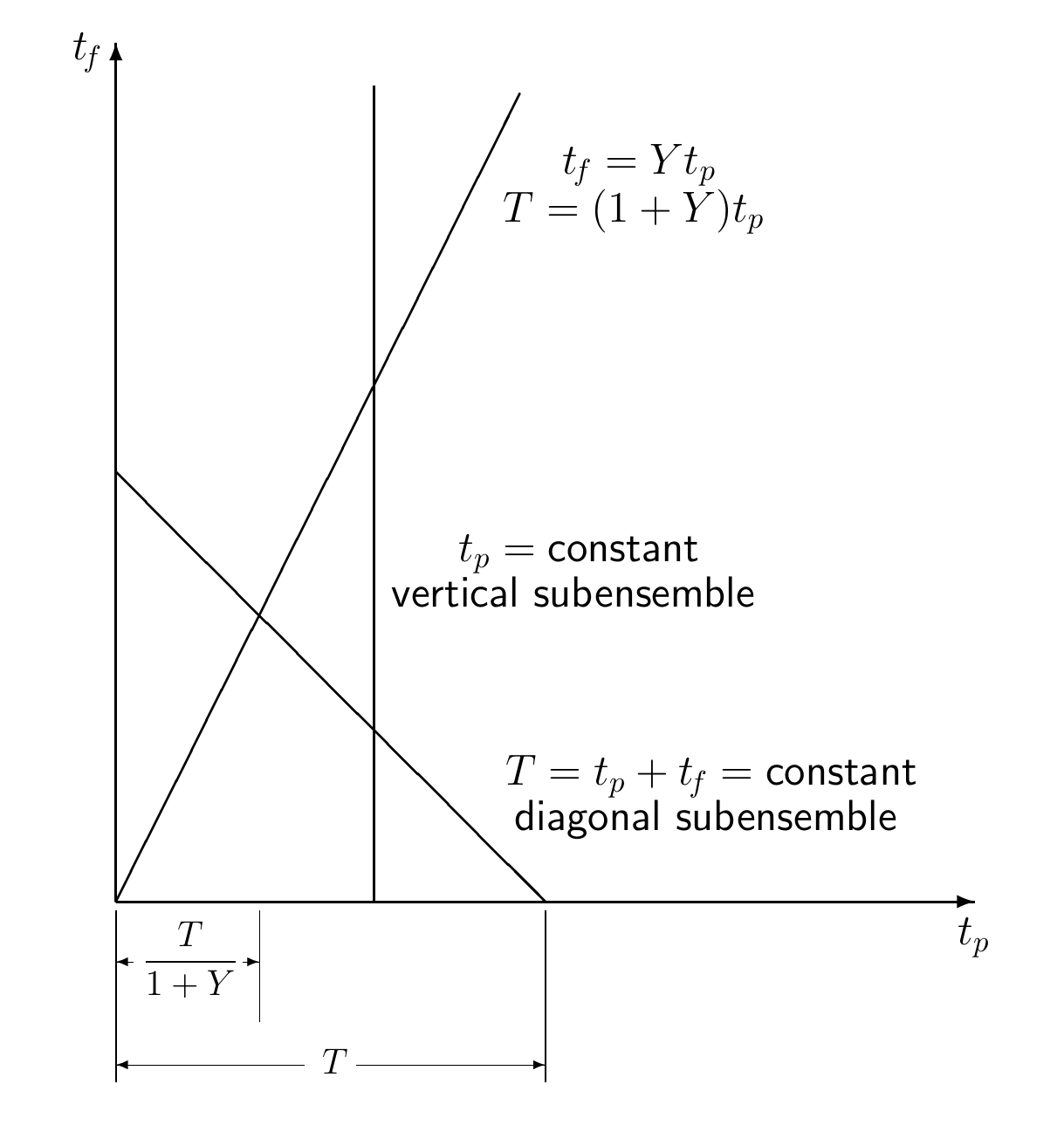}
\end{center}
\vspace{-24pt} \caption{First quadrant, on which resides the {\it
(truncated) Copernican ensemble}, which applies to a phenomenon in
progress. The truncated Copernican ensemble is obtained by lopping
off Regions~1 and~2 of the unrestricted Copernican ensemble.  The
Copernican ensemble is an idealized sample of phenomena with
uniformly random starting times $t_0$ (or present ages $\tp=-t_0$)
and with duration distributed along each vertical subensemble
according to the prior density~$w(T)$.  Since the starting time is
uniformly random, population is distributed uniformly along each
diagonal subensemble.  Gott's delta-$t$ argument is that the fraction
of population with $\tf\ge Y\tp$ within each diagonal subensemble is,
by the elementary geometry illustrated in the figure, $1/(1+Y)$.
This fraction being the same for each diagonal subensemble, it also
applies to the entire Copernican ensemble, giving Gott's
rule~(\protect\ref{eq:Grule}). The content of Gott's rule is the
trivial statement that a fraction $X$ of the members in the
Copernican ensemble have an age less than a fraction $X$ of their
eventual duration.  This trivial statement does not authorize any
prediction of future duration based on present age because the
present age is unknown.  Once the present age is known, predictions
of future duration are made within the vertical subensemble
corresponding to the observed age and thus are governed by the prior
density $w(T)$, but with those durations ruled out by the observed
age discarded.\label{fig2}}
\end{figure}

To discuss your observation requires taking into account that you are
only interested in the situation, denoted by $I$, where you find the
phenomenon to be in progress.  Imposing this condition requires you
to lop off the regions of the unrestricted Copernican ensemble that
correspond to the phenomenon not having begun or having finished
[Regions~1 and~2 of Fig.~\ref{fig1}]. This leaves the {\it
(truncated) Copernican ensemble\/} depicted in Fig.~\ref{fig2}, which
occupies the first quadrant of the $\tp$-$\tf$ plane.  The
probability density for the truncated Copernican ensemble is given by
\begin{equation}
p(\tp,\tf|I)={w(\tp+\tf)\over\overline T}\;,\quad\mbox{$\tp\ge0$, $\tf\ge0$,}
\label{eq:joint}
\end{equation}
where $\overline T$ is a normalization constant equal to the mean
value of the total duration with respect to~$w(T)$.  Truncating the
unrestricted Copernican ensemble also truncates the unrestricted
diagonal and vertical subensembles.  In the following, the
designation ``truncated'' is often omitted; an undesignated ensemble
is always the truncated one.

A {\it diagonal subensemble\/} lives on a diagonal line of constant
$T$.  Along the diagonal line, population is distributed uniformly,
and the total population is weighted by $Tw(T)$.  The Copernican
principle is the statement that the population within each diagonal
subensemble is distributed uniformly, with no bias toward the past or
the future, as is expressed by the fact that the joint
density~(\ref{eq:joint}) depends only on $T$.  A {\it vertical
subensemble\/} lives on a line of constant $t_p$; it has the same
population density as the corresponding unrestricted vertical
ensemble, except that durations that correspond to the phenomenon's
having already finished, $T<\tp$ ($\tf<0$), are not part of the
ensemble and have no population.  The Copernican ensemble is an
idealization of a sample of phenomena in progress, with random
starting times and durations distributed according to the prior
density $w(T)$.

Now suppose you ask for the probability that $\tf\ge Y\tp$ for a
phenomenon selected from the truncated Copernican ensemble.  Within
each diagonal subensemble this probability is given by the fraction
of the length of the diagonal line that lies above the line
$\tf=Y\tp$ shown in Fig.~\ref{fig2}.  This ratio, from elementary
geometry, is $1/(1+Y)$, and this ratio is the delta-$t$ argument.  Since
this fraction is the same for all the diagonal subensembles, it gives
the probability that $\tf\ge Y\tp$ within the entire truncated
Copernican ensemble.  The result is Gott's rule~(\ref{eq:Grule}),
written here as
\begin{equation}
P(\tf\ge Y\tp|I)={1\over1+Y}\;.
\end{equation}
Notice that this probability is independent of the prior density
$w(T)$; it is wholly determined by the time-translation symmetry of
the Copernican ensemble.  The rule is particularly easy to understand
for the case $Y=1$: half the members of the Copernican ensemble lie
above (below) the line $\tf=\tp$ and thus have a future duration that
is greater than (less than) their present age.  We conclude that
Gott's rule {\it is\/} a universal expression of the Copernican
principle for a phenomenon drawn from the entire Copernican ensemble,
i.e., for a phenomenon known to be in progress, but whose present age
is unknown.

Gott's rule as a universal expression of the Copernican principle has
precisely the content that a fraction $X$ of the members in the
Copernican ensemble have an age less than a fraction $X$ of their
eventual duration.  This trivial conclusion is what the Copernican
principle tells you: you know a phenomenon is in progress, but you
know neither when it started nor when it will end, so you judge
yourself equally likely to be at any point in the phenomenon's life.
This trivial conclusion is of very little interest, because the
present age being unknown, the rule has no predictive power. What
attracts attention to Gott's work is that he repeatedly uses his rule
in a different way, to make probabilistic predictions of the future
longevity of particular phenomena whose present age is known.

\begin{figure}
\begin{center}
\includegraphics[height=14cm]{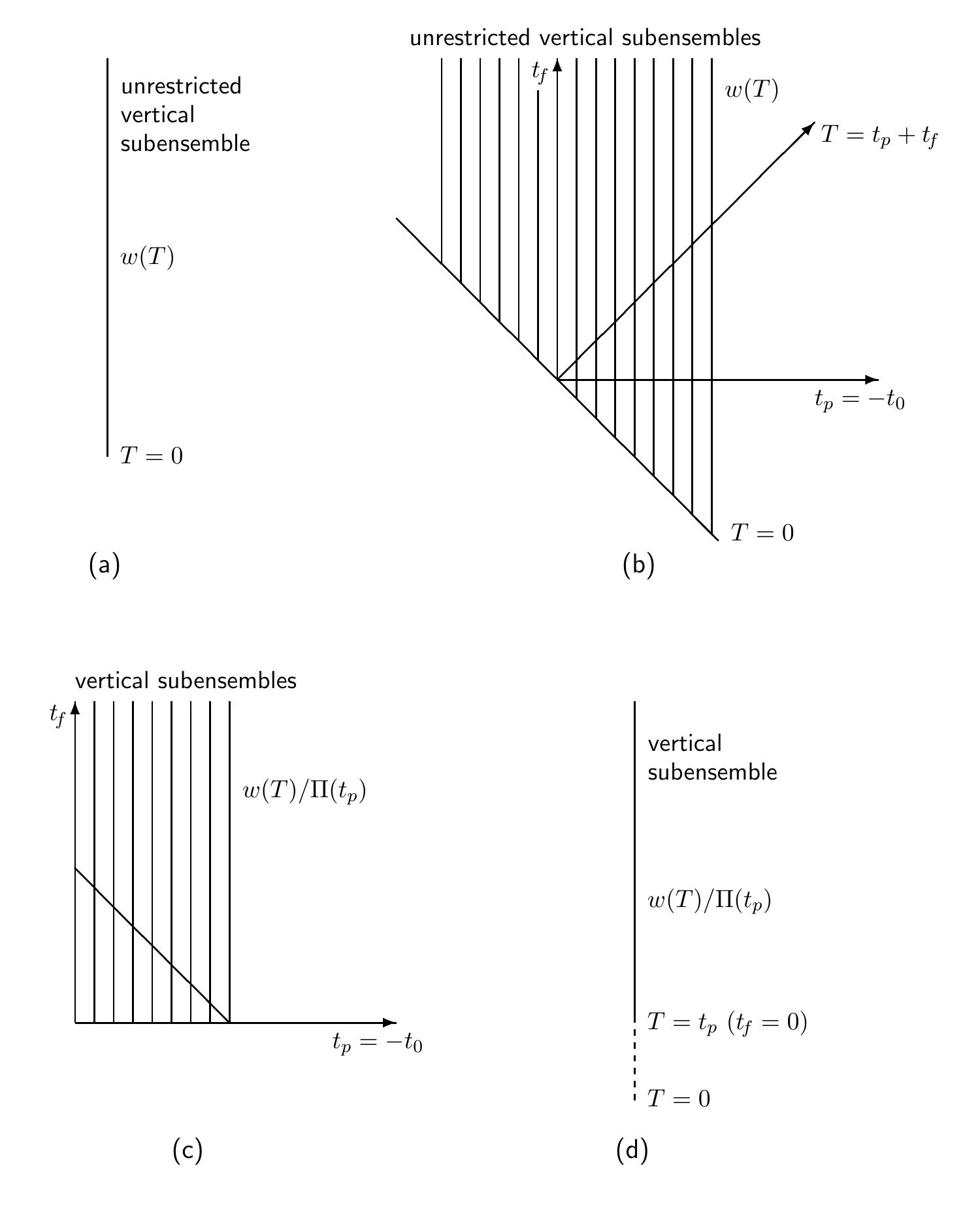}
\end{center}
\vspace{-18pt}
\caption{(a)~Unrestricted vertical subensemble, in which population
is distributed according to the prior density~$w(T)$.
(b)~Unrestricted Copernican ensemble, created from many copies of the
unrestricted vertical ensemble (fifteen copies, including the
vertical axis, are shown), each corresponding to a different starting
time $t_0=-\tp$. Gott's Copernican principle is the statement that
all starting times are equally likely. (c)~(Truncated) Copernican
ensemble, which describes phenomena in progress.  It is created by
removing from the unrestricted Copernican ensemble the regions that
correspond to phenomena not yet begun and already completed
(Regions~1 and 2 of Fig.~\ref{fig1}). In particular, each vertical
subensemble is truncated by removing the part with $T<\tp$ ($\tf<0$).
Population is distributed uniformly along the diagonal subensembles
of constant total duration~$T$, one of which is shown. (d)~Vertical
subensemble chosen by an observation of present age $\tp$.
Predictions within this vertical subensemble are governed by a
renormalized prior density, $w(T)/\Pi(\tp)$, with durations ruled out
by the observed age omitted.  Steps~(b) and~(c) of this process can
be short-circuited by going directly from~(a) to~(d).  Imagining many
copies of the unrestricted vertical ensemble, as is done in
implementing the temporal Copernican principle and thus constructing
the Copernican ensemble, or even having an approximation to the
Copernican ensemble available cannot increase your power to predict
the future duration of a phenomenon with a particular present age.
This is particularly clear in the special case of a phenomenon whose
total duration $T$ is known in advance, so that only one diagonal
subensemble is populated, say, the one shown in~(c).  At the stage of
the truncated Copernican ensemble in~(c), the present age and future
duration are strictly correlated, but randomly distributed within the
interval $[0,T]$, thus giving Gott's delta-$t$ argument.  Once you
observe the present age, however, the future duration is known and is
certainly not governed by Gott's rule.
\label{fig3}}
\end{figure}

We thus need to determine what you can say when you discover the
present age.  Your probabilistic predictions are then determined by
the distribution of population within the vertical subensemble whose
members have the observed present age.  It is clear from
Fig.~\ref{fig2} that the probability density for future duration
within this subensemble---this is the conditional probability density
for $\tf$ given $\tp$---is proportional to $w(\tp+\tf)$.  Properly
normalized, this conditional density becomes
\begin{equation}
p(\tf|\tp,I)=w(\tp+\tf)/\Pi(\tp)\;,\quad\mbox{$\tf\ge0$,}
\label{eq:cond}
\end{equation}
where the normalization constant,
\begin{equation}
\Pi(\tp)=\int_0^\infty d\tf\,w(\tp+\tf)=\int_{\tp}^\infty dT\,w(T)\;,
\end{equation}
is the survival probability, i.e., the probability for the phenomenon
to survive at least a time $\tp$.  The conditional probability
density~(\ref{eq:cond}) gives the probabilities you should use for
making predictions of future duration based on present age.  It has a
very simple interpretation: once you determine the present age, you
rule out total durations shorter than the observed age, and you use
the prior density, suitably renormalized, for total durations longer
than the observed age.  This is what you would have done had you not
bothered to introduce the Copernican ensemble, but rather worked
directly within an unrestricted vertical ensemble.$^7$

The process of constructing an unrestricted Copernican ensemble,
truncating to take account that the phenomenon is in progress, and
observing the present age is depicted in Fig.~\ref{fig3}.

One way to construct the vertical subensemble for present age $\tp$
is to select, from each diagonal subensemble with $T\ge\tp$, the
subpopulation that has age $\tp$.  That population is distributed
uniformly within the rest of each diagonal subensemble is irrelevant
to the statistics of a phenomenon drawn from a vertical subensemble.
This is why the Copernican principle has no bearing on predictions of
future duration based on present age.  Indeed, once you discover the
present age, the probability that $\tf\ge Y\tp$ is
\begin{equation}
P(\tf\ge Y\tp|\tp,I)
=\int_{Y\tp}^\infty d\tf\,p(\tf|\tp,I)
={\Pi\Bigl((1+Y)\tp\Bigr)\over \Pi(\tp)}\;.
\label{eq:rightrule}
\end{equation}
This is the predictive form of the desired probability, predictive
because it is conditioned on the present age.  It is determined
completely by the prior density and coincides with Gott's
rule~(\ref{eq:Grule}) only for a special choice of prior density,
which is identified in Sec.~\ref{sec:Bayesian} and discussed further
in Sec.~\ref{sec:conclusion}.  We conclude that Gott's rule should
not be used indiscriminately to make probabilistic predictions of
future duration based on present age.

All your prior information about a phenomenon's total duration is
incorporated in the prior density $w(T)$.  Often you can improve your
predictions of future longevity by studying a phenomenon as it
progresses, gathering information about its particular history.  In
the absence of gathering additional information, however, all
predictions about future longevity must arise from the prior density.
That Gott's rule, as it comes from the delta-$t$ argument, is
independent of the prior density is a dead give-away that it has no
predictive power.  Since any prior density can be embedded in a
Copernican ensemble, it is clear that the Copernican principle does
not restrict the prior density in any way and thus is irrelevant to
predicting future longevity.

\section{Bayesian analysis of Gott's rule}
\label{sec:Bayesian}

Gott has endorsed$^{11}$ a Bayesian derivation of his rule, which was
introduced by Buch$^{12}$ in the only technical comment {\it
Nature\/} has published on Gott's original article.  The input to
Buch's analysis is the prior density $w(T)$ and the assertion that
given the duration~$T$, present age $\tp$ is uniformly distributed
within the interval $[0,T]$:
\begin{equation}
q(\tp|T)=\cases{
1/T\;,&$0\le\tp\le T$,\cr
0\;,&$\tp>T$.}
\label{eq:qtpT}
\end{equation}
A simple application of Bayes's rule gives
\begin{equation}
q(T|\tp)={q(\tp|T)w(T)\over q(\tp)}=
\cases{
w(T)/Tq(\tp)\;,&$T\ge\tp$,\cr
0\;,&$T<\tp$,
}
\label{eq:qTtp}
\end{equation}
where
\begin{equation}
q(\tp)=\int_{\tp}^\infty dT\,{w(T)\over T}
\end{equation}
is the unconditioned probability density for present age $\tp$.  The
conditional probability that $\tf\ge Y\tp$, given $\tp$, takes the
form
\begin{equation}
Q(\tf\ge Y\tp|\tp)=
\int_{(1+Y)\tp}^\infty dT\,q(T|\tp)=
{q\Bigl((1+Y)\tp\Bigr)\over q(\tp)}\;.
\end{equation}
If you use the (unnormalizable) prior density $w(T)=1/T$, this result
reduces to Gott's rule, in a predictive form:
\begin{equation}
Q(\tf\ge Y\tp|\tp)={1\over1+Y}\;.
\end{equation}
The prior $w(T)=1/T$, called the {\it Jeffreys prior},$^{18}$ has the
unique status of being the only distribution on the interval
$[0,\infty]$ that is invariant under scale changes.  Thus this
Bayesian derivation concludes with the appealing result that Gott's
rule, as a genuinely predictive rule for future duration given
present age, follows from assuming a prior that has no built-in time
scales.

The only problem with this neat conclusion is that this Bayesian
derivation is dead wrong.  This is evident from the
posterior~(\ref{eq:qTtp}), which is not just the original prior with
excluded durations given zero probability, as in the process of
lopping off the already completed phenomena from the unrestricted
ensembles to get the truncated ensembles.  The analysis gets right
that the posterior probability is zero for durations $T<\tp$ that are
ruled out by the observation of present age $\tp$, but it doesn't use
a renormalized version of the prior density for the durations that
are still allowed, i.e., for $T\ge\tp$.  This must be wrong because
your prior density $w(T)$ already contains your entire judgment about
the future duration of the phenomenon should it survive to age $\tp$.
In the absence of getting additional information, there is nothing to
justify changing your judgment about future duration when you learn
that the phenomenon has indeed survived to age~$\tp$.

The question then is where this apparently innocuous Bayesian
analysis goes wrong.  It is not hard to determine that.  The error
lies in using the uniform conditional probability density $q(\tp|T)$
of Eq.~(\ref{eq:qtpT}) in conjunction with the prior density $w(T)$.
Within the unrestricted Copernican ensemble, where it is correct to
use $w(T)$, learning the duration $T$ tells you nothing about the
present age, as is evident from considering the unrestricted diagonal
subensemble in Fig.~\ref{fig1}. This is confirmed by a trivial
application of Bayes's rule to the uncorrelated variables $\tp$
and~$T$: $p(\tp|T)=p(t_0,T)/w(T)=\gamma$.  It is simply not
consistent with the unrestricted Copernican ensemble to use the
uniform conditional probability density~(\ref{eq:qtpT}).

The natural thing then is to try the truncated Copernican ensemble of
Fig.~\ref{fig2}, which applies once you know the phenomenon is in
progress.  Then it is correct to use a uniform conditional density
for $\tp$, i.e.,
\begin{equation}
p(\tp|T,I)=\cases{
1/T\;,&$0\le\tp\le T$,\cr
0\;,&$\tp>T$.}\;,
\end{equation}
as is evident from considering the truncated diagonal subensemble in
Fig.~\ref{fig2}, but it is not correct to use the prior
density~$w(T)$. Once you know the phenomenon is in progress, you must
weight $w(T)$ by a factor of $T$, which comes from the ``lengths'' of
the truncated diagonal subensembles being proportional to $T$.
Formally, one has
\begin{equation}
p(T|I)=\int d\tp\,d\tf\,p(\tp,\tf|I)\delta(T-\tp-\tf)={Tw(T)\over\overline T}\;.
\label{eq:pTI}
\end{equation}
The factor of $T$ here is not optional.  It is {\it required\/} once
you have decided to describe the phenomenon in terms of two temporal
variables and to impose the time-translation symmetry of the
Copernican principle on the joint probability density.  To put it
more succinctly, it is required once you decide to use an ensemble of
phenomena with random starting times.

Once one realizes that the factor of $T$ is present in $p(T|I)$,
the Bayesian inference of Eq.~(\ref{eq:qTtp}) is replaced by
\begin{equation}
p(T|\tp,I)={p(\tp|T,I)p(T|I)\over p(\tp|I)}=
\cases{
w(T)/\Pi(\tp)\;,&$T\ge\tp$,\cr
0\;,&$T<\tp$,
}
\end{equation}
since the probability density of $\tp$ is given by
\begin{equation}
p(\tp|I)=\int_0^\infty d\tf\,p(\tp,\tf|I)={\Pi(\tp)\over\overline T}\;.
\end{equation}
This correct Bayesian analysis is thus in accord with the obvious
inference of truncating the unrestricted vertical ensemble to get the
conditional probability density for $T$, given $\tp$.

Because of the additional factor of $T$ in this correct analysis, the
(unnormalizable) prior density that gives a predictive version of
Gott's rule turns out to be $w(T)=1/T^2$.  This prior density plays a
special role in this problem because it is the unique distribution on
the first quadrant of the $\tp$-$\tf$ plane that is (i) constant on
lines of constant $T$ and (ii) invariant under simultaneous scale
changes of $\tp$ and $\tf$.  Formally, with this prior, we can write
[see Eq.~(\ref{eq:rightrule})]
\begin{equation}
P(\tf\ge Y\tp|\tp,I)={1\over1+Y}\;,
\end{equation}
since $\Pi(\tp)=1/\tp$.  Thus Gott's rule, in a predictive form,
emerges from a prior $w(T)=1/T^2$ that has no time scales into the
past or future; alternatively, one can say that this predictive form
of Gott's rule arises when the probability density for $T$ within the
truncated Copernican ensemble, i.e., $p(T|I)$ of Eq.~(\ref{eq:pTI}),
is the Jeffreys prior.

\section{Conclusion}
\label{sec:conclusion}

The best way to test belief in probabilistic predictions is to offer
a bet based on those predictions.  For that purpose, I sent an e-mail
on 1999~October~21 and again on 1999~December~2 to my department's
most comprehensive e-mail alias, which included faculty, staff, and
graduate students, requesting information on pet dogs.  The responses
were compiled and checked for accuracy on 1999~December~6; a
notarized list of the 24~dogs, including each dog's name, date of
birth, breed, and caretaker, was deposited in my departmental
personnel file on 1999~December~21.  In accordance with his practice
for other phenomena, Gott would have made a prediction for each dog's
future prospects based on its age.  In particular, he would have
predicted that each dog would survive beyond twice its age with
probability $1/2$.

For the youngest and oldest dogs on the list, Gott's predictions
offered favorable opportunities for betting.  I chose to focus on the
oldest dogs, and for each of the six dogs above ten years old on the
list, I offered$^7$ to bet Gott \$1,000\,US that the dog would not
survive to twice its age on 1999~December~3.  To sweeten the pot, I
offered Gott 2:1 odds in his favor.  Gott refused the bets on the
grounds that ``I don't do bets.''$^{19}$  If he had believed his own
predictions, his expected gain would have been \$3,000\,US, and the
probability that he would have been a net loser on the six bets was
7/64=0.11.  I contacted the caretakers during May and June of 2008
and verified that all six dogs have died.  Thus, as I fully expected,
I would have won all the bets and been \$6,000\,US richer. Even with
the current reduced state of the US dollar, that would have been
enough to buy a very nice piece of Australian aboriginal art.

More revealing than Gott's blanket refusal to bet was his excuse that
his rule only applies to a random dog chosen from my sample,$^{19}$
which is another way of saying that his rule applies to a sample of
dogs drawn from the truncated Copernican ensemble, i.e., a sample
selected without regard to present age.  In discussions of the 44 New
York plays$^{\hbox{\scriptsize2--4,6}}$ and of his own
longevity,$^{2,6}$ Gott has also suggested that a fair test of the
Copernican hypothesis should involve a large sample selected without
regard to present age.  As we have seen, Gott is quite right on this
score: his rule {\it does\/} apply to a phenomenon whose present age
is unknown.  If this were all Gott claimed, however, no one would pay
attention, because the universal form of his rule, applicable when
the present age is unknown, has no predictive power.  What grabs
attention is that in case after case, Gott uses his rule to make
predictions of the future longevity of individual phenomena whose
present age is known.  In the language of this paper, Gott makes
predictions for the vertical subensembles, but only wants to bet on
the entire Copernican ensemble.

It is obvious that in a large sample of dogs selected without regard
to age, roughly half the dogs, within the inevitable statistical
fluctuations, will be in the first half of their lives, with the rest
in the second half.  This is the trivial content of Gott's Copernican
principle.  It is equally obvious that having a sample in which half
the dogs are in the first half of their lives does not imply that any
particular dog in the sample has a probability of $1/2$ to survive
beyond twice its present age.  Yet the elementary error of making
this implication underlies all of Gott's predictions.

We have seen, at the end of Sec.~\ref{sec:Bayesian}, that there is a
particular (unnormalizable) prior density, $w(T)=1/T^2$, which does
give Gott's rule in a predictive form.$^7$   Although the prior
density~$1/T^2$ does not appear in any of Gott's publications, it has
a special status in that it is the unique prior density that makes
the Copernican probability $p(\tp,\tf|I)$ invariant under
simultaneous rescaling of the past and the future.  Use of this prior
is the only license for Gott's predictions.  When you can't identify
any time scales, Gott's rule is your best bet for making predictions
of future duration based on present age.

For most phenomena, including many that Gott discusses, especially
those involving human institutions and creations, it is easy to
identify important time scales.$^7$  Although it is often difficult
to incorporate these time scales into a prior probability, it is
always a good idea to try.  This having been said, it is usually the
case that formulating prior information precisely is of less value
than observing a phenomenon as it progresses, since readily available
current information is more cogent than prior information for
predicting the future.

Although there is little love lost between White Sox fans and fans of
the Chicago Cubs, I like to think that {\sl New York Times\/} writer
Jim Glanz, having experienced a Sox World Series win in his lifetime,
sympathizes with the plight of Cubs fans, who haven't seen a World
Series title since 1908.  Gott would predict, with 95\% confidence,
that they won't win a Series in the next three years, but will win
one before 5868.  Perhaps more to the point, he would predict with
probability $1/2$ that they won't bring home a title in the next
99~years.  We are immediately skeptical of Gott's prediction.  For
example, giving each of the 30 clubs an equal chance each year sets
the probability of a 99-year drought at $(29/30)^{99}=0.035$.  It's
not that this is the ``right'' way to calculate the probability, but
it does show that a reasonable assumption gives quite a different
answer from Gott's rule.

The reason Gott's prediction for the Cubs is so unreasonable is that
there are readily identifiable time scales---the length of a typical
player's career, the turnover in owners and management, etc.---that
are well short of 99~years and suggest that the Cubs might get their
act together much sooner.  Indeed, as of June~21, they have the best
record in North American baseball and are leading the National League
Central division.  Still, Cubs fans know to keep some pessimism in
reserve.

Suppose a fan at a Cubs game at Wrigley Field in Chicago got up and
announced to great fanfare that half the people at the game were in
the first half of their life.  Everyone would yawn (except perhaps
the technically sophisticated, who might wonder about whether the
attendees are a representative sample of all ages, although a ball
game is probably not a bad sample in this regard).

Suppose, however, that the fan marched up to parents holding a
one-month-old infant and proclaimed, ``Gott says your baby has a
2.5\% chance of dying before tomorrow's game,'' or informed the
60-year-old next to him, ``Gott says you have a 50\% chance of living
to 120.''  Both these predictions would garner attention, as
applications of Gott's rule often do.  The parents would probably
call security and ask that the fan be removed.  The 60-year-old might
reply, with the ingrained pessimism of Cubs fans, ``God only knows,
but maybe if I lived to 120, I could see the Cubs win a Series.'' His
seatmate would pour cold water on that: ``Don't get your hopes up.
Gott gives, and Gott takes away.  You might live to 120, but Gott
says there's only a 38\% chance the Cubs will win the Series by
then.  There's only a 50\% chance they'll win before you're 160.''

Gott's rule makes absurd predictions for human longevity and other
human activities because there are readily identifiable time scales,
the most obvious of which is the average human life span, that render
application of his rule entirely inappropriate.  If he continues to
believe his rule makes nontrivial, universal predictions for the
future duration of individual phenomena, it's time he took some bets.

\vspace{9pt}
\hspace{1truein}\hrulefill\hspace{1truein}
\vspace{9pt}

$^1$\,J.~R. Gott~III, ``Implications of the Copernican principle for
our future prospects,'' {\sl Nature\/} {\bf 363}, 315 (1993).

\vspace{4pt}
$^2$\,J.~R. Gott~III, ``Our future in the Universe,'' in
{\sl Clusters, Lensing, and the Future of the Universe}, Astronomical
Society of the Pacific Conference Series, Vol.~88, edited by
V.~Trimble and A.~Reisenegger (Astronomical Society of the Pacific,
San Francisco, 1996), p.~140.

\vspace{4pt}
$^3$\,J.~R. Gott~III, ``A grim reckoning,'' {\sl New Scientist\/}
{\bf 156}\,(No.~2108), 36 (1997 November~15).

\vspace{4pt}
$^4$\,J.~R. Gott~III, ``The Copernican principle and human survivability,''
in {\sl Human Survivability in the 21st Century}, Transactions of the
Royal Society of Canada, Series~VI, Vol.~IX, edited by D.~M.~Hayme
(University of Toronto Press, Toronto, 1999), p.~131.

\vspace{4pt}
$^5$\,J.~R. Gott~III, ``Colonies in space; Will we plant colonies
beyond the Earth before it is too late?" {\sl New Scientist\/}
{\bf 195}\,(No.~2620), 51 (2007 September~8).

\vspace{4pt}
$^6$\,J.~R. Gott~III, {\sl Time Travel in Einstein's Universe\/}
(Houghton Mifflin, Boston, 2001), Chap.~5.

\vspace{4pt}
$^7$\,C.~M. Caves, ``Predicting future duration from present age:
A critical assessment,'' {\sl Contemporary Physics\/} {\bf 41}, 143
(2000).

\vspace{4pt}
$^8$\,T.~Ferris, ``How to predict everything: Has the physicist
J.~Richard Gott really found a way?'' {\sl The New Yorker\/}
{\bf 75}\,(18), 35 (1999 July~12).

\vspace{4pt}
$^9$\,J. Tierney, ``A survival imperative for space colonization,''
{\sl The New York Times\/} (2007 July~17).

\vspace{4pt}
$^{10}$\,J. Glanz, ``Point, counterpoint and the duration of everything,''
{\sl The New York Times\/} (2000 February~8).

\vspace{4pt}
$^{11}$\,J.~R. Gott~III, ``Future prospects discussed---Gott replies,''
{\sl Nature\/} {\bf 368}, 108 (1994).

\vspace{4pt}
$^{12}$\,P.~Buch, ``Future prospects discussed,'' {\sl Nature\/}
{\bf 368}, 107 (1994).

\vspace{4pt}
$^{13}$\,A.~Ledford, P.~Marriott, and M. Crowder, ``Lifetime prediction from
only present age: Fact or fiction?'' {\sl Physics Letters~A\/} {\bf 280},
309 (2001).

\vspace{4pt}
$^{14}$\,K.~D. Olum, ``The doomsday argument and the number of possible
observers,'' {\sl Philosophical Quarterly\/} {\bf 52}, 164 (2002).

\vspace{4pt}
$^{15}$\,E.~Sober, ``An empirical critique of two versions of the doomsday
argument---Gott's line and Leslie's wedge,'' {\sl Synthese\/} {\bf 135},
415 (2003).

\vspace{4pt}
$^{16}$\,L.~Bass, ``How to predict everything: Nostradamus in the role of
Copernicus,'' {\sl Reports of Mathematical Physics\/} {\bf 57}, 13 (2006).

\vspace{4pt}
$^{17}$\,B.~Monton and B.~Kierland, ``How to predict future duration from
present age,'' {\sl Philosophical Quarterly\/} {\bf 56}, 16 (2006).

\vspace{4pt}
$^{18}$\,E.~T.~Jaynes, {\sl Probability Theory: The Logic of
Science}, edited by G.~L. Bretthorst (Cambridge University Press,
Cambridge, England, 2003), Chap.~12.

\vspace{4pt}
$^{19}$\,``Life, longevity, and a \$6\,000 bet,'' {\sl
Physics World} (2000 February~11), {\tt http://physicsworld\urldot
com\urlslash cws\urlslash article\urlslash news\urlslash 2890}.
Quotes from Gott in this short article are based on a long and
informative document entitled ``Random observations and the
Copernican Principle,'' which Gott posted to PhysicsWeb early in
2000~February.  As of this writing, I have been unable to find this
long document anywhere on the web, but I have a copy that can be made
available on request.

\vspace{9pt}
\hspace{1truein}\hrulefill\hspace{1truein}
\vspace{9pt}

{\it Author's note:} This paper was originally submitted to {\sl
Nature\/} on 2000 April~3 and was summarily rejected on the grounds
that {\sl Nature\/} had already published sufficient technical
comment$^{12}$ on Gott's original paper.  Then I forgot about it,
though it's not clear to me now why I didn't post it to the preprint
archive.  That was probably just pique, to which I was more subject
then than now.  It's just as well, because the current version is, I
think, considerably improved.  Three circumstances prompted me to
revive the paper: (i)~my 2007--08 sabbatical at the University of
Queensland, which has given me the gift of time; (ii)~John Tierney's
recent {\sl New York Times\/} article,$^9$ which showed that Gott's
predictions still have the power to fascinate; and (iii)~a
conversation with B.~J. Brewer of the University of Sydney, which
indicated that there would be interest in a simpler explanation of my
{\sl Contemporary Physics\/} article.$^7$  In preparing the current
version, I expanded the discussion of the delta-$t$ argument with the
aim of making it as simple and airtight as possible, incorporated a
discussion of Bayesian derivations of Gott's rule, updated the
references, and gathered information about the six dogs' ultimate
fates.

\end{document}